\documentclass[acmtog]{acmart}
\usepackage{duckuments}
\usepackage{algorithmic}
\usepackage{algorithm}
\DeclareMathOperator*{\argmax}{arg\,max}

\usepackage{siunitx}
\usepackage{bm}
\usepackage{booktabs}
\usepackage{duckuments}
\usepackage{url}
\usepackage{array}
\usepackage{multirow}
\usepackage{multicol}

\AtBeginDocument{%
  }

\setcopyright{acmlicensed}
\copyrightyear{2025}
\acmYear{2025}
\acmDOI{XXXXXXX.XXXXXXX}
\acmConference[Conference acronym 'XX]{Make sure to enter the correct conference title from your rights confirmation email}{2025}{XXXX, XXXX}
\acmISBN{978-1-4503-XXXX-X/2018/06}

\acmSubmissionID{XXX}

\begin{document}

\title{Computational Design and Fabrication of Protective Foam}

\author{Tsukasa Fukusato}
\affiliation{%
  \institution{Waseda University}
  \city{Tokyo}
  \country{Japan}}
\email{tsukasafukusato@waseda.jp}

\author{Naoki Kita}
\affiliation{%
  \institution{Shinshu University}
  \city{Nagano}
  \country{Japan}}
\email{nkita@cs.shinshu-u.ac.jp}

\renewcommand{\shortauthors}{Fukusato and Kita}

\begin{abstract}
This paper proposes a method to design protective foam for packaging 3D objects. Users first load a 3D object and define a block-based design space by setting the block resolution and the size of each block. The system then constructs a block map in the space using depth textures of the input object, separates the map into two regions, and outputs the regions as foams. The proposed method is fast and stable, allowing the user to interactively make protective foams. The generated foam is a height field in each direction, so the foams can easily be fabricated using various materials, such as LEGO blocks, sponge with slits, glass, and wood. 
This paper shows some examples of fabrication results to demonstrate the robustness of our system.
In addition, we conducted a user study and confirmed that our system is effective for manually designing protective foams envisioned by users.
\end{abstract}

\ccsdesc[500]{Computing methodologies~Shape modeling}
\ccsdesc[500]{Computing methodologies~Graphics systems and interfaces}

\keywords{Protective Foam, Depth Texture, Region Growing, User Interface}


\maketitle

\section{INTRODUCTION}
\label{sec:intro}
\begin{figure}[t]
 \centering
 \includegraphics[width=\linewidth]{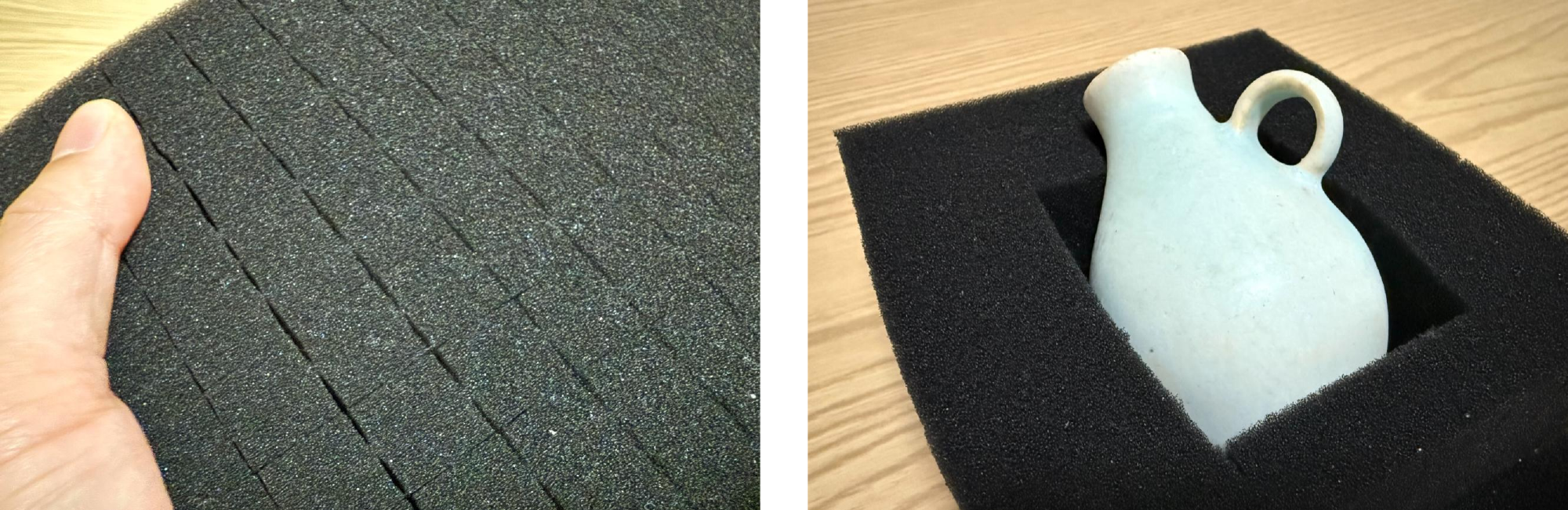}
 \caption{An example of protective foam made of sponge material. This foam is from TRUSCO~(\url{https://www.trusco.co.jp/en/)}.}
 \label{fig:example}
\end{figure}

Digital fabrication technologies, such as 3D printers and CNC milling machines, have attracted attention; however, if the strength of generated 3D objects (e.g., plastic models and figures) is not sufficient, they may be damaged during transportation. 
Under these circumstances, ``protective foam'' is often used. Protective foam is a plastic, polystyrene, or sponge material that fills the gap between 3D objects and their prepared transportation case (e.g., cardboard box), as shown in \autoref{fig:example}. 
However, it is not trivial to shape the material for such packaging as the gap between the 3D object and the transportation case must be minimized while imagining the 3D object’s rotation angle, which requires special skills and time-consuming manual work. Here, our final goal is to support the designing of such foams.

In this paper, we propose a method to generate protective foams for packaging 3D objects. The design space is a 3D grid that is compatible with actual protective foam, and we are limited to protective foams that can be taken in and out from the $x$-axis direction. For our experiment, we implemented a prototype system to interactively design protective foams referring to standard modeling tools. Our method is simple enough to implement, and it can be incorporated into existing systems easily. In summary,
our work contains the following key contributions:

\begin{itemize}
\item A novel method to make fabricable foams for input objects in real-time.
\item A user study of the protective foam design, demonstrating the benefit of the proposed system.
\end{itemize}

\section{RELATED WORK}
\label{sec:relatedWork}
In the field of digital fabrication, some authors have explored to reinterpret the manufacturing process of mold casting and make reusable casting molds. For example, ShapeCast~\cite{lyons2024shape} enables users to design plaster molds from a single 2D profile. Although this system can simplify the mold generation process, it is unsuitable for complex object casting. 
Some approaches start by decomposing an input 3D model into several regions corresponding to mold pieces~\cite{alderighi2018metamolds, nakashima2018corecavity, stein2019interactive, alderighi2022state}. However, since these approaches aim to reproduce the details of the 3D model as similar as possible through press processing, it is necessary to (somewhat) deform the input 3D model~\cite{sorkine2007asap}. That is, these are unsuitable for creating protective foams for manufactured 3D objects (that cannot be deformed). 

PackMold~\cite{kita2024packmolds} enables users to design molds suitable for desktop thermoforming; specifically, it is used to create packaging such as blister packs. Although shape characteristics in the objects' height field can be taken into account without the deformation process, its optimization takes time (about $20$ to $150$ seconds), making interactive design difficult. In addition, its design space is on the height field, which makes it unsuitable for filling all the gaps between the 3D object and the transportation case when packing the obect inside the case. Taking a somewhat similar approach, Igarashi et al.~\cite{igarashi2009interactive} propose a tool to design customized covers from a given 3D object. 
This method allows users to manually specify seam lines and make 2D patterns, but fabricable shapes are limited to the convex hull of the input model. That is, their design space is on the convex hull, and handling more complex 3D objects remains difficult.
The main purpose of protective foam is not to reproduce fine details but to simply fill in the gaps between 3D objects and a transportation case to protect the objects from damage caused by external forces. In these situations, the design space should cover the space between the transportation case and the input object instead of the object surface.

Therefore, by referring to actual foams (e.g., sponge material with cuts at regular intervals, as shown in \autoref{fig:example}), and existing voxel-based fabrication research~\cite{bacher2014spin, larsson2020tsugite, prevost2013make}, we defined a block-based design space and consider a method to interactively design protective foams from 3D models.

\section{USER INTERFACE} 
\label{sec:ui}
\begin{figure}[t]
 \centering
 \includegraphics[width=\linewidth]{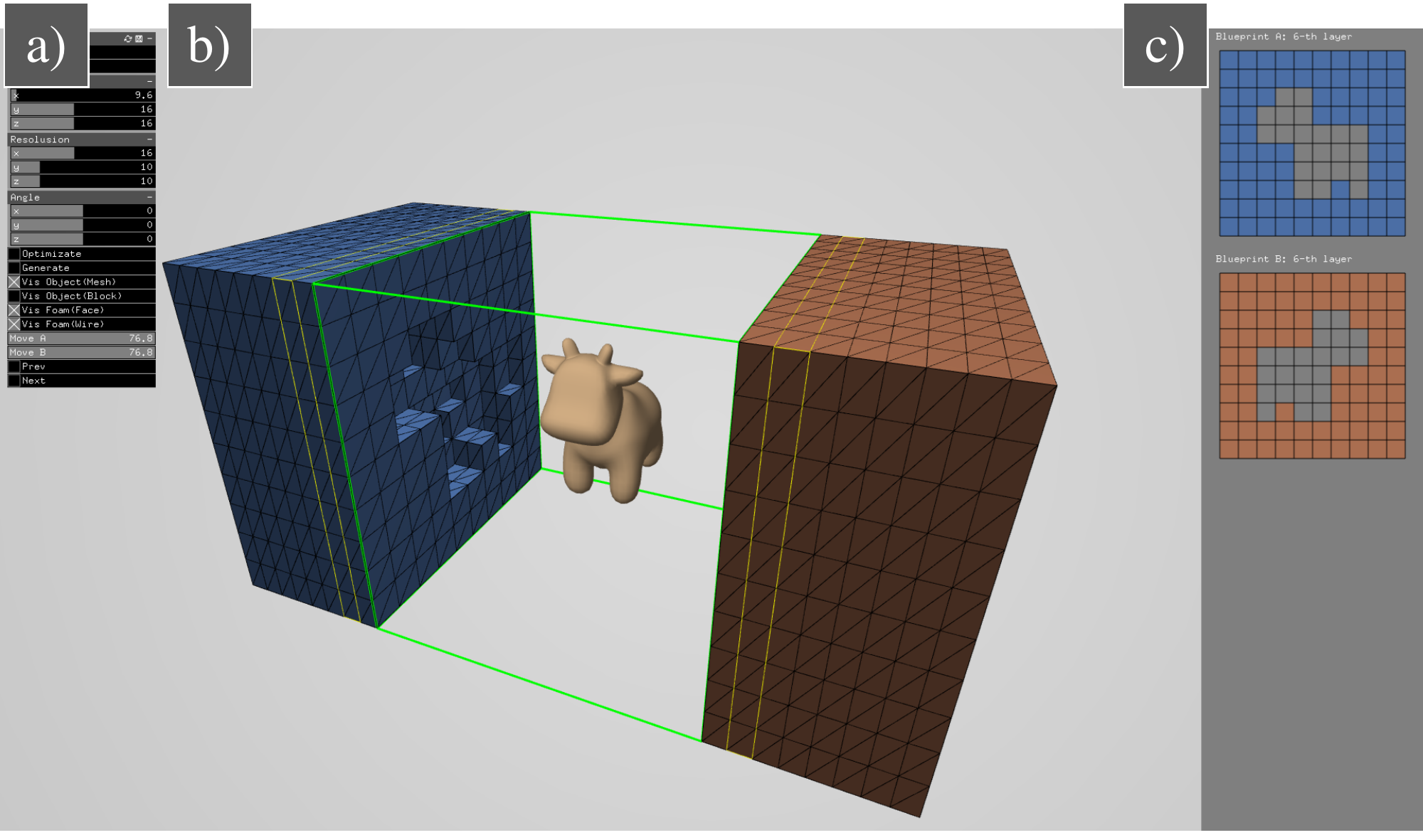}
 \caption{Screenshot of our system that identifies (a)~the tool buttons/sliders, (b)~the 3D window displaying the protective foams for the input object, and (c)~the 2D window showing slice representations of protective foams perpendicular to $x$-axis.}
 \label{fig:ui}
\end{figure}

We build on standard 3D modeling tools and add several functions to design protective foams. Our prototype system was implemented on a $64$-bit Windows $11$ laptop (Intel\textcircled{\scriptsize R}Core$^{TM}$ i$7$-$1165$G$7$ CPU@$2.80$GHz and $32.00$GB RAM) using standard OpenGL and GLSL, as shown in \autoref{fig:ui}.

\begin{figure*}[t]
 \centering
 \includegraphics[width=\linewidth]{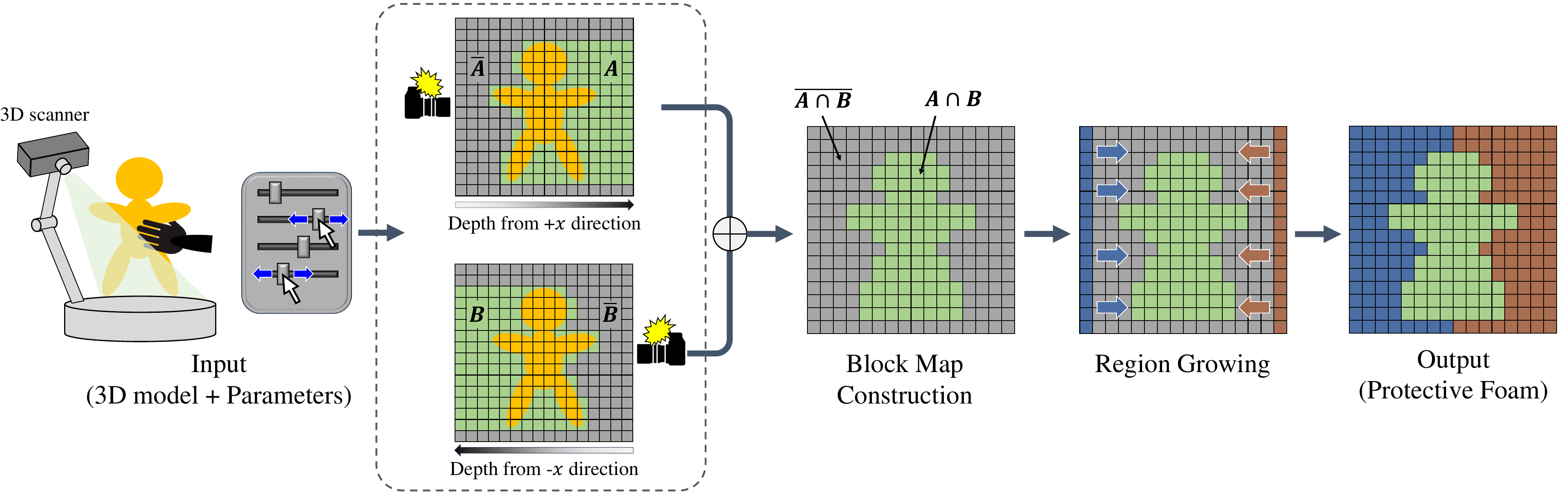}
 \caption{Overview of protective foam design. The user first load (or scan) a 3D object and specifies the parameters of the block-based design space. From these inputs, the system computes the 3D protective foams (blue and orange) and its slices.}
 \label{fig:overview}
\end{figure*}

\autoref{fig:overview} provides an overview of our protective foam design. The user first loads a 3D object scanned using a commercial 3D scanner into our system. Note that the scanned data (point cloud) needs to be converted into a polygon soup using an existing meshing tool~\cite{fukusato2020interactive} or software~\cite{cignoni2008meshlab}. Other operations are described below.

\subsection{Block Setting Function}
The user sets the block resolution and the size of each block (i.e., width~$\times$~height~$\times$~depth). The system then computes a design space where we can generate protective foams and visualizes it on the screen, as shown in \autoref{fig:ui}~(green). The initial size of block piece is \SI{15}{[mm]}$\times$\SI{15}{[mm]}$\times$\SI{22}{[mm]}, which is referring to the TRUSCO's sponge foam, and the initial block resolution is empirically set to $30\times18\times18$.

\subsection{Generation Function}
When the user clicks on the ``generate'' button, the system renders $+x$ and $-x$ depth textures of the input object from two cameras according to the size of the design space while performing hidden surface removal. The camera positions are fixed outside the design space, and their directions are from the camera position to the origin. 
Next, the system resizes the depth textures to the design space resolution so that the resized textures, named $A$ and $B$, include all pixel depths that overlap or touch the visible surface of the input object. Based on the resized textures, we construct a block map~$BM$ as follows:
\begin{equation}
BM = \overline{A \cap B}
\label{eq:foam}
\end{equation}
where $\cap$ and $-$ are the intersection operator and the complement operator, respectively. Note that to check the intersection blocks, depth texture B is mirror-flipped for alignment. Compared to the conventional voxelization of 3D models~\cite{schwarz2010fast}, our approach with two depth textures makes it difficult to represent regions that are not visible from $+x$ and $-x$ direction, but is suitable for our problem setting, which is to design protective foams that allow users to take 3D models in and out of a transportation case. 

Next, we divide the obtained block map~$BM$ into two regions (blue and orange) by using a region growing algorithm from $+x$ and $-x$ direction, and output the divided regions as protective foams. Our protective foams are height fields, so they can be easily made by various systems, such as 3D printers (without support structures), CNC milling machines, and mold casting~\cite{valkeneers2019stackmold, lyons2024shape}. 

The generation process is sufficiently fast to return immediate feedback (in less than a second) when the user edits the parameters of blocks or rotates the 3D object. Our system can output the results in common 3D model formats (i.e., .ply and .stl). 


\subsection{Rotation Angle Setting Function}
\label{subsec:angle}
\begin{figure}[t]
 \centering
 \includegraphics[width=\linewidth]{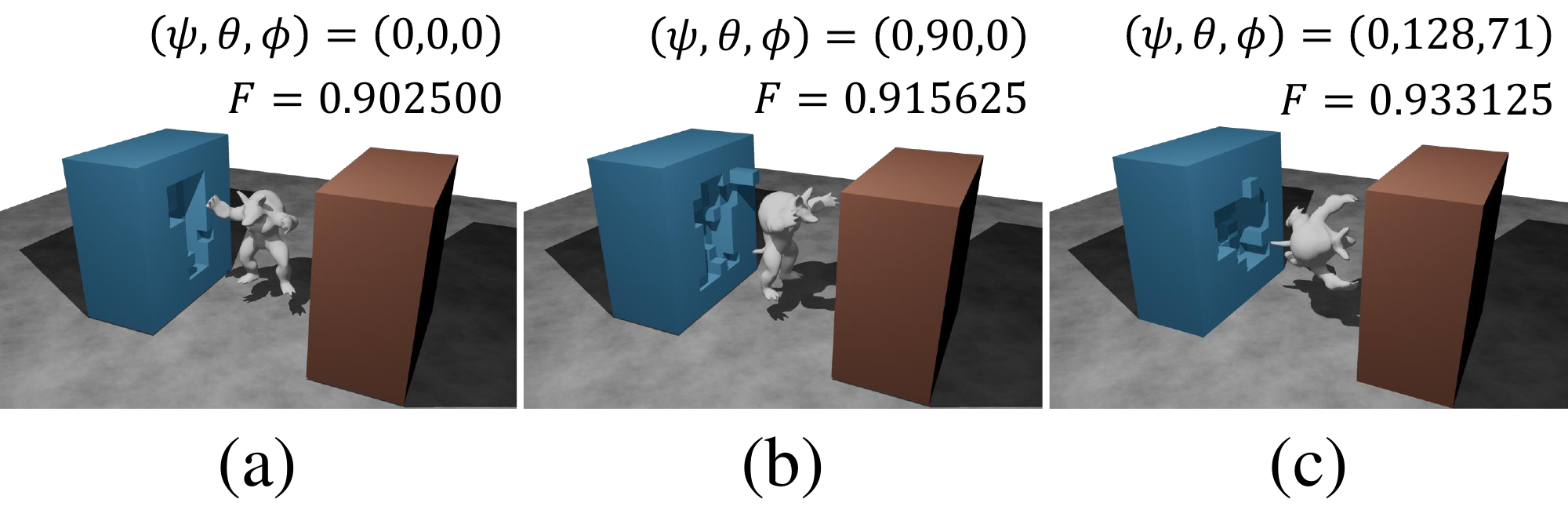}
 \caption{The initialization of the rotation angle. (a, b) randomly assigning results, and (c) our heuristic.}
 \label{fig:angle}
\end{figure}

By adjusting the ``angle'' sliders, the user can freely change the rotation angle of a 3D object based on Euler angles \{$\psi$, $\theta$, $\phi$\}.
However, depending on the rotation angle, our foams may make gap area between 3D objects and the generated foam since its foams are made by two depth textures only. Large gap areas can cause 3D object damage. Thus, we implemented a function to initialize the rotation angle that minimizes the gap areas when clicking the ``optimize'' button. 
Here, the main objective is not to fully-automatically select an optimal rotation angle, but rather to provide a good starting point for manual editing. For this, inspired by Prevost et al.~\cite{prevost2013make}, we use a simple heuristic that prefers angles with large numbers of block map, namely
\begin{equation}
\argmax_{\psi, \theta, \phi} F (\psi, \theta, \phi)
\label{eq:angle}
\end{equation}

\noindent
where $F(\psi, \theta, \phi)$ is the total volume of $BM$ generated with the given rotation angle. Note that we normalized the volume using the size of the design space. To maximize this reward function, we employ a standard greedy algorithm, which is easy to compute and enables us to set the angle sliders' values almost instantly, as shown in \autoref{fig:angle}.

\subsection{Visualization Function}
\begin{figure}[t]
 \centering
 \includegraphics[width=\linewidth]{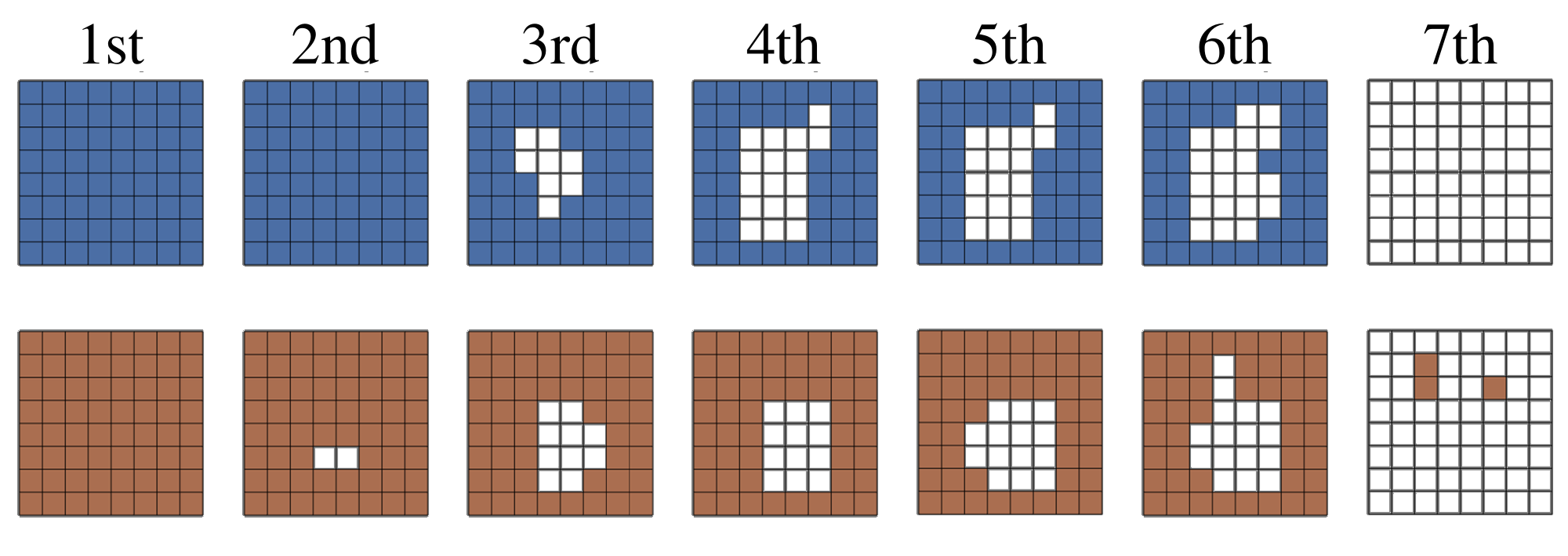}
 \caption{An example of visualizing the slices of the generated foams for the Stanford Bunny.}
 \label{fig:vis}
\end{figure}

After generating the foam, the user can freely turn the generated protective foams around by performing a mouse-drag operation and take the 3D object in and out of the foams using ``move'' sliders on the modeling panel. In addition, our system has a simple function to visualize the slices of the generated protective foams perpendicular to $x$-axis, as shown in \autoref{fig:vis}. This function is simple but useful for designing multi-stage foams like StackMold~\cite{valkeneers2019stackmold}.

\section{RESULTS}
\label{sec:result}
\begin{figure*}[t]
 \centering
 \includegraphics[width=\linewidth]{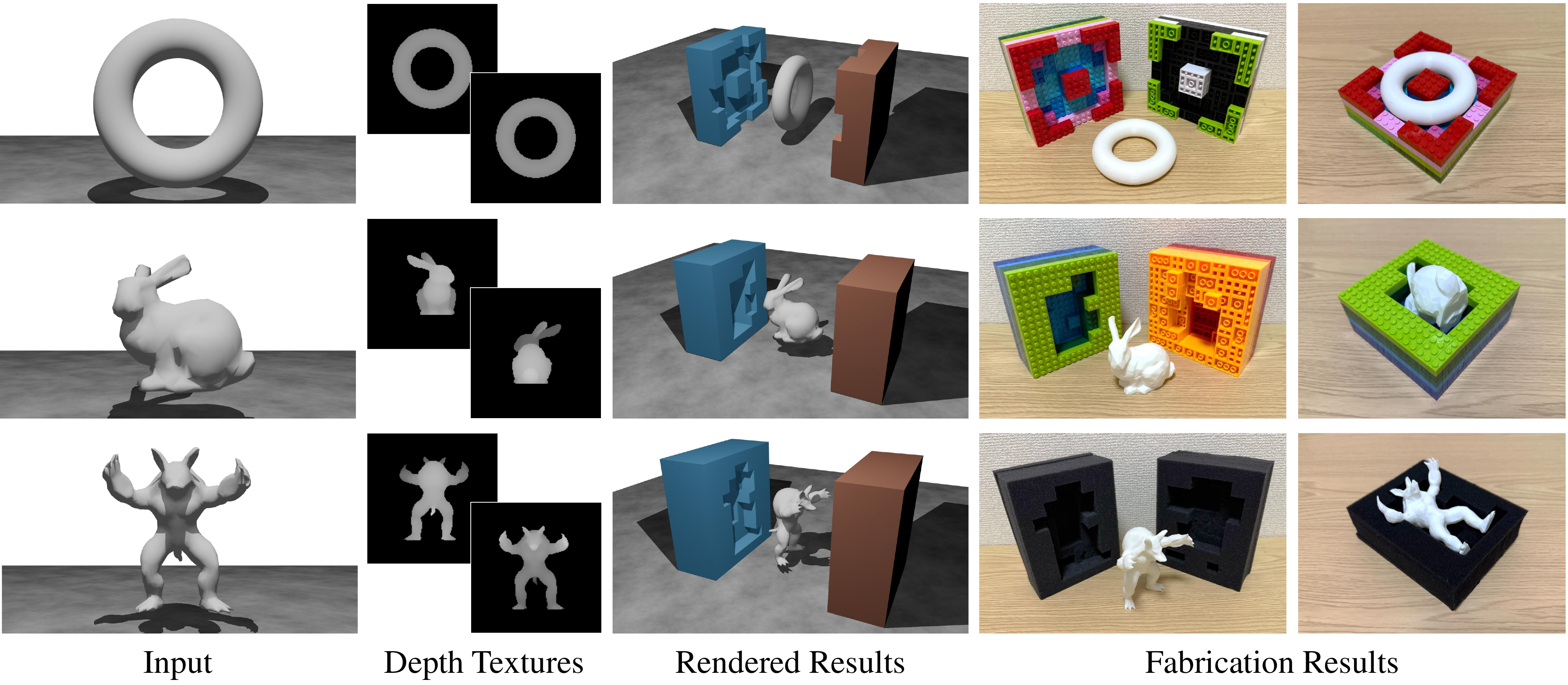}
\caption{Examples of fabricated foams with LEGO blocks and sponge materials from (Top)~Torus, (Middle)~Stanford Bunny, and (Bottom)~Stanford Armadillo. Each resolution of the design space is $8\times8\times8$, $12\times8\times8$, and $6\times10\times10$, respectively.}
 \label{fig:fabrication}
\end{figure*}

\autoref{fig:fabrication} shows the results of fabricated protective foam using LEGO blocks and sponge materials. 
Note that Luo's method~\cite{luo2015legolization} was used to optimize the LEGO block combination. 
From the results, we confirmed all materials can provide sufficient protection since the gaps between the target object and the designed foam are small enough. 
Of course, these foams can be fabricated with other materials and methods, such as plastic via a 3D printer, silicone, glass (e.g., glasswork that puts toys inside glass), and wood (e.g., building blocks).

\begin{figure}[t]
 \centering
 \includegraphics[width=\linewidth]{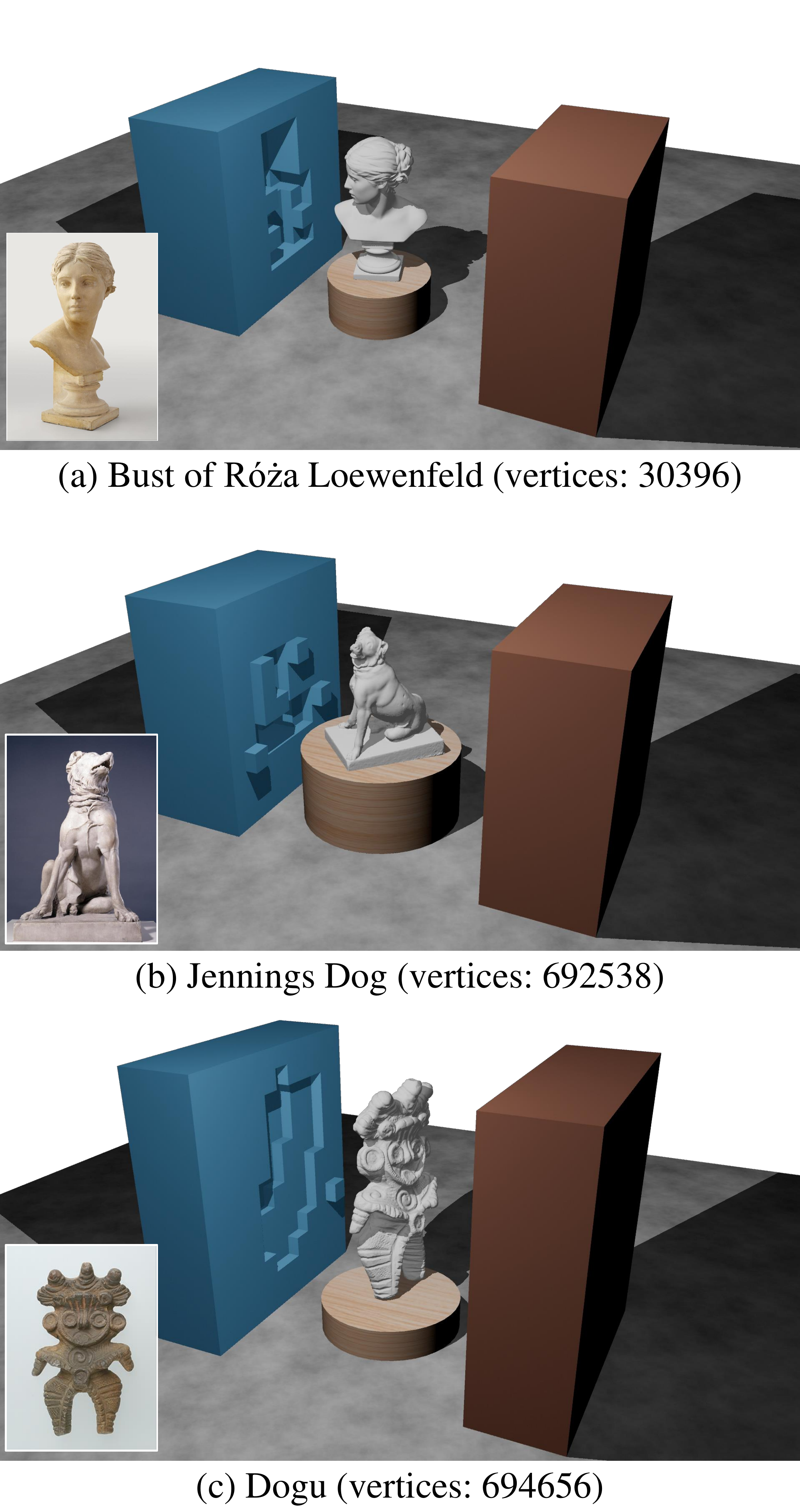}
 \caption{Examples of protective foams from scanned statues/figures. The input objects have been made available on Sketchfab~(\protect\url{https://sketchfab.com}) by the Małopolska Virtual Museums, the British Museum, and the Osaka Museum of History.}
 \label{fig:scan}
\end{figure}

In addition, we investigated usage scenarios in which the proposed system would be utilized by users who frequently use protective foams. One museum group commented, ``\textit{when transporting statues/figures for special exhibitions and their conservation, museum staff often prepare many sponges/polystyrene materials and manually make protective foams to prevent the statues from being damaged. Of course, the foam shape must be designed to securely fit statues/figures, as they are fragile. But it is difficult to repeatedly check the fitting due to the strength of the statues/figures, and the staff often fail to make foams. As a result, they have to fix it with cellophane tape}.'' From this comment, our system can be useful when packaging 3D objects in art museums, where trial and error is difficult, as shown in \autoref{fig:scan}.

\autoref{tab:ctime} shows the computation time for generating protective foams at different resolutions for the 3D models. From these results, our system, even with a single thread, is fast enough to interactively design protective forms for high-res models such as scanned statues~/~figures. The main reason is that, unlike standard fabrication methods, our system does not apply shape processing algorithms into the input models and only makes low-res depth textures. It is thought that it might speed up previous fabrication methods by using depth textures.

\begin{table}[t]
\centering
 \caption{The computation time of protective foam generation using single-threaded computing.}
 \begin{tabular}{c|c|c|c}
     \toprule
     Model & 
     \#V & 
     Resolution & 
     Time~[ms]\\
     \midrule
     \multirow{5}{*}{Torus} &
     \multirow{5}{*}{7200}& 
     $10\!\times\!10\!\times\!10$& 
     12.0 \\
     &&
     $15\!\times\!15\!\times\!15$& 
     16.8 \\
     &&
     $20\!\times\!20\!\times\!20$& 
     40.8 \\
     &&
     $25\!\times\!25\!\times\!25$& 
     62.2 \\
     &&
     $30\!\times\!30\!\times\!30$& 
     110.4 \\
     \midrule
     \multirow{5}{*}{Stanford Bunny} &
     \multirow{5}{*}{502}& 
     $10\!\times\!10\!\times\!10$& 
     12.6 \\
     &&
     $15\!\times\!15\!\times\!15$& 
     16.8 \\
     &&
     $20\!\times\!20\!\times\!20$& 
     40.8 \\
     &&
     $25\!\times\!25\!\times\!25$& 
     73.0 \\
     &&
     $30\!\times\!30\!\times\!30$& 
     114.8 \\
     \midrule
     \multirow{5}{*}{Armadillo} &
     \multirow{5}{*}{1502}& 
     $10\!\times\!10\!\times\!10$& 
     12.6 \\
     &&
     $15\!\times\!15\!\times\!15$& 
     17.2 \\
     &&
     $20\!\times\!20\!\times\!20$& 
     45.0 \\
     &&
     $25\!\times\!25\!\times\!25$& 
     70.2 \\
     &&
     $30\!\times\!30\!\times\!30$& 
     133.8 \\
     \midrule
     \multirow{5}{*}{\begin{tabular}{c}Bust of R{\`o}{\.z}a\\ Loewenfeld\end{tabular}} &
     \multirow{5}{*}{30396}& 
     $10\!\times\!10\!\times\!10$& 
     14.8 \\
     &&
     $15\!\times\!15\!\times\!15$& 
     20.6 \\
     &&
     $20\!\times\!20\!\times\!20$& 
     45.4 \\
     &&
     $25\!\times\!25\!\times\!25$& 
     70.6 \\
     &&
     $30\!\times\!30\!\times\!30$& 
     117.8 \\
     \midrule
     \multirow{5}{*}{Jennings Dog} &
     \multirow{5}{*}{692538}& 
     $10\!\times\!10\!\times\!10$& 
     21.6 \\
     &&
     $15\!\times\!15\!\times\!15$& 
     25.6 \\
     &&
     $20\!\times\!20\!\times\!20$& 
     49.0 \\
     &&
     $25\!\times\!25\!\times\!25$& 
     72.6 \\
     &&
     $30\!\times\!30\!\times\!30$& 
     135.4 \\
     \midrule
     \multirow{5}{*}{Dogu} &
     \multirow{5}{*}{694656}& 
     $10\!\times\!10\!\times\!10$& 
     42.6 \\
     &&
     $15\!\times\!15\!\times\!15$& 
     49.6 \\
     &&
     $20\!\times\!20\!\times\!20$& 
     79.4 \\
     &&
     $25\!\times\!25\!\times\!25$& 
     103.8 \\
     &&
     $30\!\times\!30\!\times\!30$& 
     143.2 \\
     \bottomrule 
 \end{tabular}
\label{tab:ctime}
\end{table}
\section{USER STUDY}
\label{sec:userstudy}
We conducted a user study to gather feedback regarding our system from participants.

\begin{table*}[t]
\centering
 \caption{Results of the post-experiment questionnaire.
$\Uparrow$ indicates higher scores are better, $\Downarrow$ for the other case.}
 \begin{tabular}{c|l|c|c}
     \toprule
     \# & Question & Mean & SD \\
     \midrule
     1 & I think that I would like to use the modeling tool frequently. $\Uparrow$ & 4.50 & 0.67 \\
     2 & I found the modeling tool unnecessarily complex. $\Downarrow$& 1.30 & 0.46 \\
     3 & I thought the modeling tool was easy to use. $\Uparrow$& 4.90 & 0.30 \\
     4 & I think that I would need the support of a technical person to be able to use this modeling tool. $\Downarrow$& 2.50 & 0.92 \\    
     5 & I found that the various functions in this modeling tool were well integrated. $\Uparrow$ & 4.70 & 0.64 \\    
     6 & I thought there was too much inconsistency in the modeling tool. $\Downarrow$& 1.30 & 0.46 \\    
     7 & I would imagine that most people would learn to use the modeling tool very quickly. $\Uparrow$& 4.40 & 0.45 \\    
     8 & I found this modeling tool very cumbersome to use. $\Downarrow$& 1.40 & 0.49 \\    
     9 & I felt very confident using this modeling tool. $\Uparrow$& 3.80 & 0.98 \\    
     10 & I needed to learn a lot of things before I could get going with this modeling tool. $\Downarrow$ & 2.10 & 0.94 \\
     \midrule
     11 & Rating of the design support for protective foams. $\Uparrow$& 4.10 & 0.94 \\    
     12 & Rating of the quality satisfaction with the designed foams. $\Uparrow$& 4.70 & 0.46 \\  
    \bottomrule
 \end{tabular}
\label{tab:sus}
\end{table*}

\subsection{Procedure}
We invited 10 participants (P1, $\dots$, and P10) aged 20--50 years ($Avg. = 29.14$, $SD = 14.12$). Each participant was asked to fill out a form asking about their experience in fabricating 2D/3D objects. 
P1 (1 male) had extensive experience creating 3D models with Blender ($>$ 4 years), creating robots ($>$ 5 years), and fabricating objects with an XYZ printer as a hobby.
P2--8 (5 male and 2 female) had prior knowledge of 3D modeling with the Autodesk Maya and Blender. 
P2 also had professional game programming experience with Unity. 
P9--10 (2 female) were experienced users of fabrication systems, such as computerized sewing machines ($>$ 3 years), but had no 3D modeling experience.

First, we gave the participants a brief overview of our system. The instructor gave a step-by-step tutorial to familiarize the participants with the foam design framework. After the overview, they could smoothly design protective foams. 
Next, we provided them with some 3D objects and asked them to keep designing protective foams for 3D objects until they were satisfied.
After completing the foam design, each participant was asked to fill out a post-experiment form. The form includes (Q1--Q10)~a standard system usability scale (SUS) to verify the usefulness of our system~\cite{brooke1996sus}, (Q11)~a scale for rating the design support for protective foams, (Q12)~a scale for rating the quality satisfaction of the designed protective foam, and two free comment questions about the good and bad points of the proposed system (optional). Note that Q11 and Q12 were answered on a 5-point Likert scale (1: strongly dissatisfied to 5: strongly satisfied). When the participants were filling out the questionnaire, we also conducted a casual interview with them to talk about their impressions about our system.

\subsection{Observations and User Feedback}
\autoref{tab:sus} shows the post-experiment questionnaire results, giving the mean values (Mean) and standard deviations (SD). From the Q1--10 results, the final SUS score is $84.25$, which is calculated by averaging each participant’s SUS score. The standard SUS score is 68, and our score of $84.25$ is regarded as excellent and has the grading scale of Grade A. In addition, according to the Q11 \& Q12 results, we confirmed that the participants were satisfied with the design support and generated results ($> 4$). 

Next, we summarized the participants' comments regarding our system below.

\begin{itemize}
\item P1: \textit{It was pretty cool since given the user-specified parameters, the system can generate natural-looking foams in real-time.}
\item P5: \textit{The system's operation screen is very simple and easy to use.}
\item P10: \textit{It's useful to be able to quickly make blueprints of protective foams when transporting relatively-expensive objects like cameras and musical instruments urgently.}
\end{itemize}

Overall, the participants reported that our system was straightforward to use and useful. A possible reason is that our system enables users to make fabricable foams in real-time, and can be used as a base tool. In the future, it might be interesting to explore the possibility of incorporating other functions into our tool. There were some requests from participants to add functions as follows:

\begin{itemize}
\item P6: \textit{Once the user presses the generate button, I would like the foam shape to automatically update when manipulating the sliders/buttons.}
\item P7\&8: \textit{In the fabrication step, I thought it would be easier to understand if the depth values of blocks were visualized instead of slices of the generated foams.}
\item P9: \textit{I want a function to automatically calculate the minimum size of transportation case while minimizing the gaps for reducing transportation cost and labor.}
\end{itemize}

According to these comments, we found that some participants identified several issues with the current implementation, but found these not to be serious problems. In the future, we plan to improve the user experience with further engineering.

\section{LIMITATIONS AND FUTURE WORK}
\label{sec:limitation}
Our current system does not handle situations in which multiple parts, for example, a camera body and lens, are packaged. For practical purposes, it might be interesting to consider constraints for collision handling between two objects to pack multiple objects into one transportation case. 
Additionally, in the current implementation, the case shape is limited to cuboid boxes. It might be better to allow users to add a function to interactively pack 3D objects into a case with an arbitrary shape, such as one with smooth surfaces. 

The present paper focuses only on filling the gaps between the object and the case, so it is still difficult to consider pressure distribution during packaging. We will explore the best design that can protect all fragile regions. We also plan to explore the possibility of extending our idea (i.e., the design support for the space around 3D objects) to other design tasks, such as connectors~\cite{koyama2015auto}.

\section{CONCLUSION}
\label{sec:conclusion}
This paper has presented a method to design protective foams for 3D objects. 
We define a block map from two depth textures of the input object and extract two foams using a region growing algorithm. The system can easily visualize and export the designed foams, allowing users to fabricate them with existing fabrication process. In addition, through a user study, we confirmed that the proposed system was highly
appreciated by novices/amateurs with 3D modeling or fabrication experience. Hence, we believe that our method will be a new step toward the acceleration of research in computational fabrication research.


\bibliographystyle{ACM-Reference-Format}
\bibliography{sample-base}

\end{document}